\documentclass[pra,showpacs,superscriptaddress,10pt,twocolumn]{revtex4-2}
\usepackage{epsfig}
\usepackage{epstopdf}
\usepackage{delarray}
\usepackage{amsmath, amssymb}
\usepackage{bm}
\usepackage{graphicx}
\usepackage{color} 
\usepackage{amssymb}
\usepackage[figuresright]{rotating}
\usepackage{float} 
\usepackage{lipsum}
\usepackage{epstopdf} 
\usepackage{amsmath}

\usepackage[section]{placeins}

\begin{document}

\title{Lagrangian Drifter Path Identification and Prediction: SINDy vs Neural ODE}

\author{Cihan Bay\i nd\i r}
\affiliation{\.{I}stanbul Technical University, Engineering Faculty, 34469 Maslak, \.{I}stanbul, Turkey.}
\email{cbayindir@itu.edu.tr}

\author{Fatih Ozaydin}
\affiliation{Tokyo International University, Institute for International Strategy, 4-42-31 Higashi-Ikebukuro, Toshima-ku, Tokyo 170-0013, Japan.\\  
Nanoelectronics Research Center, Kosuyolu Mah., Lambaci Sok., Kosuyolu Sit., No: 9E/3 Kadikoy, \.{I}stanbul, Turkey.}
\email{fatih@tiu.ac.jp}

\author{Azmi Ali Altintas}
\affiliation{\.{I}stanbul University, Science Faculty, Physics Department, 34116 Vezneciler, \.{I}stanbul, Turkey.}
\email{azmi.altintas@istanbul.edu.tr}

\author{Tayyibe Erişti}
\affiliation{\.{I}stanbul Technical University, Engineering Faculty, 34469 Maslak, \.{I}stanbul, Turkey.}
\email{eristi24@itu.edu.tr}

\author{Ali Rıza Alan}
\affiliation{\.{I}stanbul Technical University, Engineering Faculty, 34469 Maslak, \.{I}stanbul, Turkey.}
\email{alan21@itu.edu.tr}

\begin{abstract}
In this study, we investigate the performance of the sparse identification of nonlinear dynamics (SINDy) algorithm and the neural ordinary differential equations (ODEs) in identification of the underlying mechanisms of open ocean Lagrangian drifter hydrodynamics with possible applications in coastal and port hydrodynamic processes. With this motivation we employ two different Lagrangian drifter datasets acquired by National Oceanic and Atmospheric Administration (NOAA)'s surface buoys with proper World Meteorological Organization (WMO) numbers. In the SINDy approach, the primary goal is to identify the drifter paths of buoys using ordinary differential equation sets with a minimal number of sparse coefficients. In the neural ODE approach, the goal is to identify the derivative of the hidden state of a neural network (NN). Using the acquired data, we examine the applicability of the SINDy and the neural ODE algorithms in identification of the drifter trajectories comparatively. We propose that while both of the algorithms may give acceptable results for open ocean, the SINDy-based algorithmic approach can predict the Lagrangian drifter paths  more accurately and consistently at least for the datasets investigated and parameters selected. A discussion of our findings with potential applications in search and rescue missions in the open ocean, their limitations and applicability are also presented.
\end{abstract}
\maketitle



\section{Introduction}
Modeling hydrodynamic impacts such as waves, current, turbulence, and Lagrangian circulation observed in coastal and harbor zones, understanding the instruments of different occasions like coastal disintegration and the spread of contamination, as well as making forecasts, requires an exertion to comprehend the complexity of the marine environment. One of the most commonly used techniques to analyze these effects is Lagrangian drifters \cite{Davis}. Extracting Eulerian velocity fields from a cluster of drifter trajectories requires a large number of drifters and data \cite{Garraffo2, Ohlmann1}. Also, the Eulerian velocity field analysis usually focuses on longer time scales \cite{Ohlmann2}. Whereas Lagrangian assessments are mostly based on drifter trajectories and their statistics, making them a more popular choice.

Throughout the globe, researchers constantly use many different Lagrangian drifters to measure oceanographic parameters with different spatial and temporal resolutions \cite{Elipot, Lin, Liu}. Some of these parameters include but are not limited to surf-zone currents \cite{MacMahan, Spydell}, headland rip currents \cite{McCarroll}, oil spills \cite{Jordi} and Eulerian or Lagrangian statistics \cite{Garraffo1, Griffa, LaCasce}. Model simulated drifter trajectories of these and similar phenomena can be directly compared with independent drifter observations \cite{Thompson, Barron} to ensure the validity of the modeling. With increasing coverage of remote and in situ data, Lagrangian drifter trajectories are becoming more useful in assessing the performance of numerical ocean circulation models \cite{Ohlmann2, Toner, vanSebille}.

Lagrangian trajectory identification and evolution is a subject of many different studies \cite{Mitarai, Jordi, Chen, McClean, LaCasce}. In these studies, it is found that the Lagragian drifter prediction errors tend to grow as a function of time. The error growth rate is found to be proportional to the square root of the velocity variance \cite{Mitarai}. Also it is observed that model performance assessment should consider different dynamical flow regimes separately \cite{Mitarai, Jordi, Chen}. Thus, ocean circulation patterns should be known as a priori \cite{Mitarai, Jordi, Chen}. Additionally, the actual ocean circulation may turn out to be quite different than that inferred from climatological mean patterns \cite{Mitarai, Jordi, Chen}.

Also, some algorithmic approaches for identification and the prediction of the Lagrangian drifter trajectories are proposed in the literature. The Lagrangian estimating the travel time and the most likely path from Lagrangian drifters is studied in \cite{Omalley}.
A nonlinear trajectory diagnostic approach depending on the Ornstein–Uhlenbeck (OU) process that underlines the importance of uncertainty is built to construct phase portrait maps of Lagragian drifters is studied in \cite{ChenN}. Augmented-state Lagrangian data assimilation method using the local ensemble Kalman transform filter is investigated in \cite{Sun}, where a realistic regional ocean data assimilation system is used. Seven commonly used methods for  mapping out the dominant Lagrangian coherent structures, namely i) finite-time Lyapunov exponent, ii) trajectory path length, iii) trajectory correlation dimension, iv) trajectory encounter volume, v) Lagrangian-averaged vorticity deviation, vi) dilation, and vii) spectral clustering techniques were applied to surface drifters in the Gulf of Mexico in \cite{Rypina}.

In this paper, we approach the Lagrangian path identification and prediction problem from a different perspective. We employ the SINDy and the neural ODE algorithms for these identification and prediction purposes and present a comparison of the successes of these algorithms in tackling these problems. SINDy algorithm is recently being proposed for Lagrangian drifter path identification and prediction by our research group in studies \cite{Teristi, BaySINDYinfus, UluAlan,alan2023predictability}. Our findings in those studies are promising for drifter path identification. However, to our best knowledge neural ODE algorithms have never been tested before with those purposes, and no intercomparison of these two algorithms is provided in the literature. By employing two different Lagrangian drifter data sets, we analyze the applicability of the SINDY vs neural ODEs, discuss our findings and their limitations.

\section{Methodology}

\subsection{Study Area and Lagrangian Drifter Data}
Two different Lagrangian drifter data sets used in this study. Both of the data sets are acquired from the National Oceanic and Atmospheric Administration (NOAA)'s Physical Oceanography Division (PhOD)' web portal which can be accessed at {\url{https://www.aoml.noaa.gov/phod/gdp/real-time_data.php}}.  

The first data set is recorded by the buoy with WMO identification number of $1601747$, which is a surface drifter providing the latitudes (Lat) and longitudes (Lon) of its trajectory. This drifter is referred to as Drifter 1 throughout this study. The initial and final dates and times of data acquisition used in our study are $2024-01-03T \ 11:00:00Z$ and $2024-03-10T \ 06:00:00Z$, respectively. The start and end coordinates of the drifter trajectory are [Lat, Lon]: $-50.1406, 29.5426$ and [Lat, Lon]: $-43.7982,46.6184$, respectively. The sampling time of Drifter 1 is set to be $1$ hr. The Drifter 1 trajectory plotted on a geomap can be seen in Figs.~\ref{f:1},~\ref{f:2} and~\ref{f:3} in the following section.

The second data set is recorded by another buoy with WMO identification number of $5501700$ which is referred as Drifter 2 from this point forward. As Drifter 1, Drifter 2 is also a surface drifter providing the latitudes (Lat) and longitudes (Lon) of its trajectory. The initial and final dates and times of data acquisition used in our study are $2024-01-30T \ 16:00:00Z$ and $2024-02-25T \ 16:00:00Z$, respectively. The start and end coordinates of the drifter trajectory are [Lat, Lon]: $-43.2248, -166.3598$ and [Lat, Lon]: $-43.1922,-162.4752$, respectively. The sampling time of Drifter 2 is also $1$ hr. The Drifter 2 trajectory on another geomap can be seen in Figs.~\ref{f:4} and~\ref{f:5} in the following section.

\subsection{Review of the Sparse Identification of Nonlinear Dynamics (SINDy) Algorithm}
Being one of the algorithms employed for the Lagrangian drifter trajectory identification and prediction, a brief review of the sparse identification of nonlinear dynamics (SINDy) algorithm first introduced in \cite{Brunton} is provided in this subsection. We start by considering a dynamical system in the form of 
\begin{equation} 
\frac{d {\bf{x}}(t)}{dt}={\bf{f}}(x(t))
\label{eq_1}
\end{equation}
which is capable of identifying many motion dynamics observed in the nature~\cite{Brunton}. In this formulation, ${\bf{x}}(t)$ shows the state of the system investigated at time $t$. The equation of the motion of the system considered is defined by the function ${\bf{f}}(x(t))$ \cite{Brunton}. A large set of potential candidate functions for ${\bf{f}}$ are constructed by the SINDy algorithm, which then uses an $l_1$-regularized regression which promotes sparsity to determine the dominant terms among those possible candidate functions \cite{Brunton}. For the determination of the $f$ from the data, a time series of the state ${\bf{x}}(t)$ is used as 
\begin{equation}\label{eq_2}
    \bf{X} = \left[{\bf{x}}(t_1) \ {\bf{x}}(t_2) \hdots {\bf{x}}(t_m) \right]^T				
  \end{equation}
where the terms $T$, $m$ and $n$ terms stands for the transpose operation, the number of observations in time and the dimension of state ${\bf{x}}$, respectively \cite{Brunton}. After the ${\bf{x}}$ observations are performed, the derivative matrix given as
		\begin{equation}
    \bf{\dot{X}} = \left[{\bf{\dot{x}}}(t_1)  \ {\bf{\dot{x}}}(t_2) \hdots {\bf{\dot{x}}}(t_m) \right]^T
				\label{eq_3}
  \end{equation}
can be calculated using many different simple approaches, or which can also be directly measured as well \cite{Brunton}. Then, SINDy constructs a possible set of nonlinear polynomial and trigonometric functions in the form of	
\begin{equation} 
{\bf{\Theta}}({\bf{X}})=\left[1 \  {\bf{X}} \ {\bf{X}}^2 \ \hdots {\bf{X}}^d  \ \hdots  \sin({\bf{X}}) \ \hdots \right]
				\label{eq_4}
\end{equation}
which are constructed using $\bf{X}$. In this formula, $d$ stands for the degree of polynomials. The $\sin({\bf{X}})$ term represents some candidate trigonometric functions such as $\sin({\bf{X}})$, $\sin(2{\bf{X}})$, etc. The time derivatives to the candidate nonlinear functions can be related as
	\begin{equation}
    \bf{\dot{X}} = {\bf{\Theta}}({\bf{X}}) \Xi
				\label{eq_5}
  \end{equation}
where $\Xi$ matrix contains the column vectors, $\xi_k$ \cite{Brunton}. After determination of those sparse coefficient vectors, $\xi_k$, one can write
	
	\begin{equation}
    {\dot{x}}_k = {\bf{\Theta}}({\bf{x}}) \xi_k.
				\label{eq_6}
  \end{equation}
The SINDy algorithm uses a least absolute shrinkage and selection operator (LASSO) as an $l_1$-regularized regression which promotes sparsity \cite{Brunton}. Thus, the regression problem can be formulated as
		\begin{equation}
    {\xi_k}=\underset{{{\xi}'_k}}{\textnormal{arg min}} || {\bf{\dot{X}}}_k-{\bf{\Theta}}({\bf{X}}) {{\xi}'_k} ||_{{}_2} + 
\lambda || {{\xi}'_k} ||_{{}_1}	
				\label{eq_7}
				\end{equation}
where $\lambda$ is the regularization parameter and is set to be $\lambda=0.05$ for all calculations in this work following \cite{Brunton}. The sparse $\xi_k$ vectors are identified among all possible ${\xi}'_k$ candidates \cite{Brunton}. The reader is referred to \cite{Brunton} for a more comprehensive discussion of the SINDy algorithm.

In the literature, many extensions of the SINDY algorithm are proposed. There extensions include but are not limited to SINDy with control \cite{Brunton_SINDyC}, SINDy for boundary value problems \cite{Shea} and SINDY for Lagragian identification \cite{Purnomo} where some artificial intelligence (AI) and deep learning (DL) techniques have also been implemented to enhance the retrieval of the motion dynamics via SINDy algorithm. The SINDy algorithm and some of its extensions are applied to some of the celebrated problems in hydrodynamics \cite{Fukami}. In this paper, we investigate the possible usage of the SINDy algorithm for the retrieval of the dynamics of the Lagrangian drifters similar to some of our recent work in \cite{Teristi, BaySINDYinfus, UluAlan} and compare the results obtained by the SINDy algorithm with the neural ODE algorithm summarized in the next subsection.

\subsection{Review of the Neural Ordinary Differential Equation (Neural ODE) Algorithm}
Neural ODEs treat the hidden state of a NN, $\mathbf{x}(t)$, as a continuous-time dynamical system \cite{Chen, Shampine}. The evolution of the hidden state $\mathbf{x}(t)$ is defined by the following ODE

\begin{equation}
    \frac{d\mathbf{x}(t)}{dt} = \mathbf{f}(\mathbf{x}(t), t; \theta)
		\label{eq_8}
\end{equation}
where $\mathbf{f}$ is a neural network parametrized by $\theta$ \cite{Chen, Shampine}. The initial condition is given as $\mathbf{x}(0) = \mathbf{x}_0$, allowing for the prediction of $\mathbf{x}(t)$ at any time $t$. Instead of specifying a discrete sequence of hidden layers, neural ODE parametrizes the derivative of the hidden state using a NN \cite{Chen, Shampine}. The output is computed using a black-box ODE solver \cite{Chen, Shampine}. The output is computed using a black-box ODE solver \cite{Chen, Shampine}.  These continuous-depth neural ODE models have constant memory cost and adapt their evaluation strategy according to each input \cite{Chen, Shampine}. In this paper, the neural ODE that defines the Lagrangian drifter data is solved numerically with the explicit Runge-Kutta45 algorithm of the MATLAB software. The backward pass is used in automatic differentiation to learn the trainable parameters $\theta$ by backpropagating through each operation of the ODE solver. The number of hidden states is selected to be 20 and a learning rate of 0.002 is used where an epoch consists of at least 200 batchsize iterations. Different iteration such as 700, 1000, 1500 are tested for the Lagrangian drifter path identification as illustrated below. These parameters are applied to all simulations presented in this paper. The reader is referred to studies \cite{Chen, Shampine} for a more comprehensive discussion of the neural ODE algorithms.

\section{Results and Discussion}
\begin{figure*}[t!]
	\begin{center}
		\includegraphics[width=0.8\textwidth]{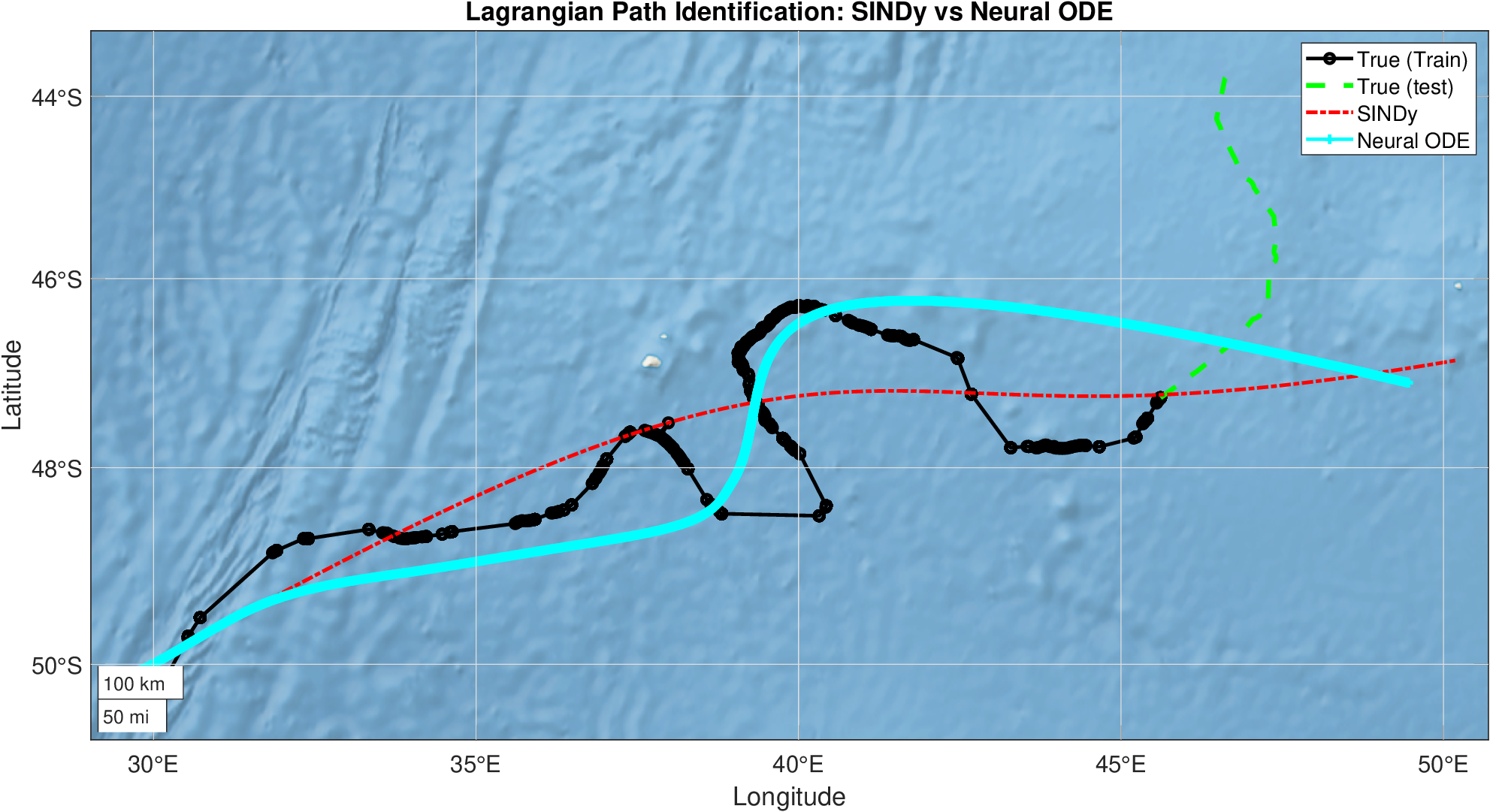}	
	\caption{\small The comparison of the SINDy using $d=3$ degree polynomials and no trigonometric terms vs neural ODE using $700$ iterations in identification and prediction of the Drifter 1 (WMO$\#$ 1601747) trajectory.}
	\label{f:1}
\end{center}
\end{figure*}

As discussed above, the data used in this study is downloaded from the NOAA website. The first data set is depicted in Fig.~\ref{f:1}, together with the identifications and predictions performed by the SINDy and neural ODE algorithm. The $75\%$ of the observational data is used as a training set, and the remaining $25\%$ is tested against prediction throughout this study. The upper limit of the relative tolerances of both of the SINDy and neural ODE retrieval algorithms is selected to be $10^{-7}$ for all calculations presented in this study. For the analysis presented in Fig.~\ref{f:1}, SINDy algorithm with $d=3$ degree polynomial order is used without any trigonometric terms. The number of iterations is selected to be 700 for neural ODE algorithm considering computation times that particularly becomes an important parameter in the case of acute marine disasters.

\begin{figure*}[t!]
\begin{center}
   \includegraphics[width=0.8\textwidth]{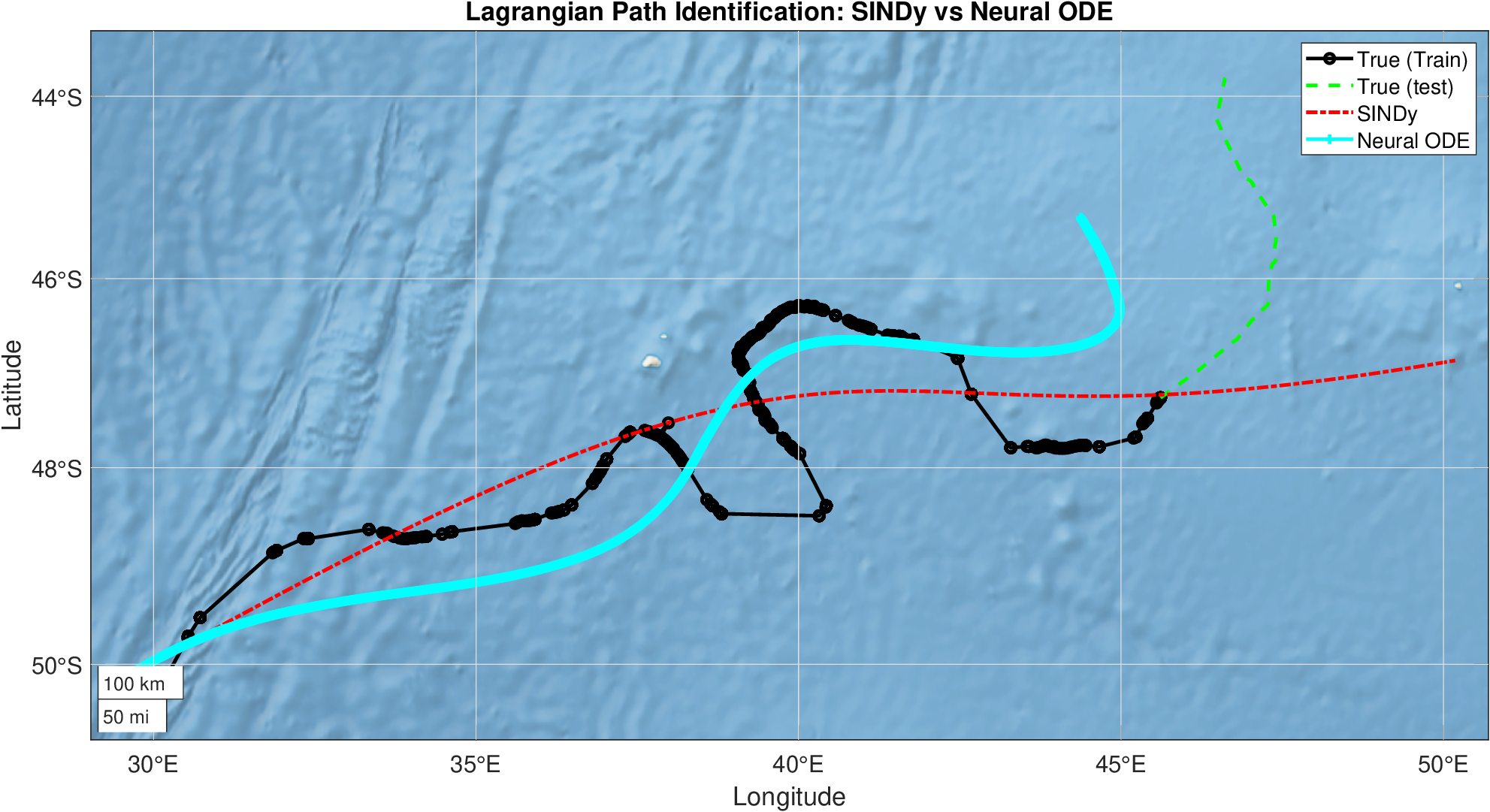}
\caption{\small The comparison of the SINDy using $d=3$ degree polynomials and no trigonometric terms vs neural ODE using $1500$ iterations in identification and prediction of the Drifter 1 (WMO$\#$ 1601747) trajectory.}
  \label{f:2}
\end{center}
\end{figure*}

All distance calculations between two points of the Lagrangian trajectories on the ocean surface is calculated by utilizing the World Geodetic System of 1984 (WGS84) formula of the MATLAB software. Accordingly, the WGS84 distance between the endpoints of the actual trajectory vs the trajectory identified by SINDY with parameters given above is calculated as $4.4132\textnormal{x}10^5$ m. Whereas, the distance between the endpoints of the actual trajectory vs the trajectory identified by neural ODE with parameters given above
is calculated to be $4.2983\textnormal{x}10^5$ m. Although the distance between the end point of the actual observations vs neural ODE predictions is smaller than its SINDy analog, it is not possible to mention than neural ODE is a more successful Lagrangian drifter tracking algorithm as illustrated in Fig.~\ref{f:1}. As discussed below for different runs with different parameters, SINDY appears to be a more reliable tool because of its consistency in retrieving the same trajectory for different runs. Whereas, the convergence and the consistency of the neural ODE is not guaranteed. This result is illustrated in Fig.~\ref{f:2}.

\begin{figure*}[t!]
\begin{center}
   \includegraphics[width=0.8\textwidth]{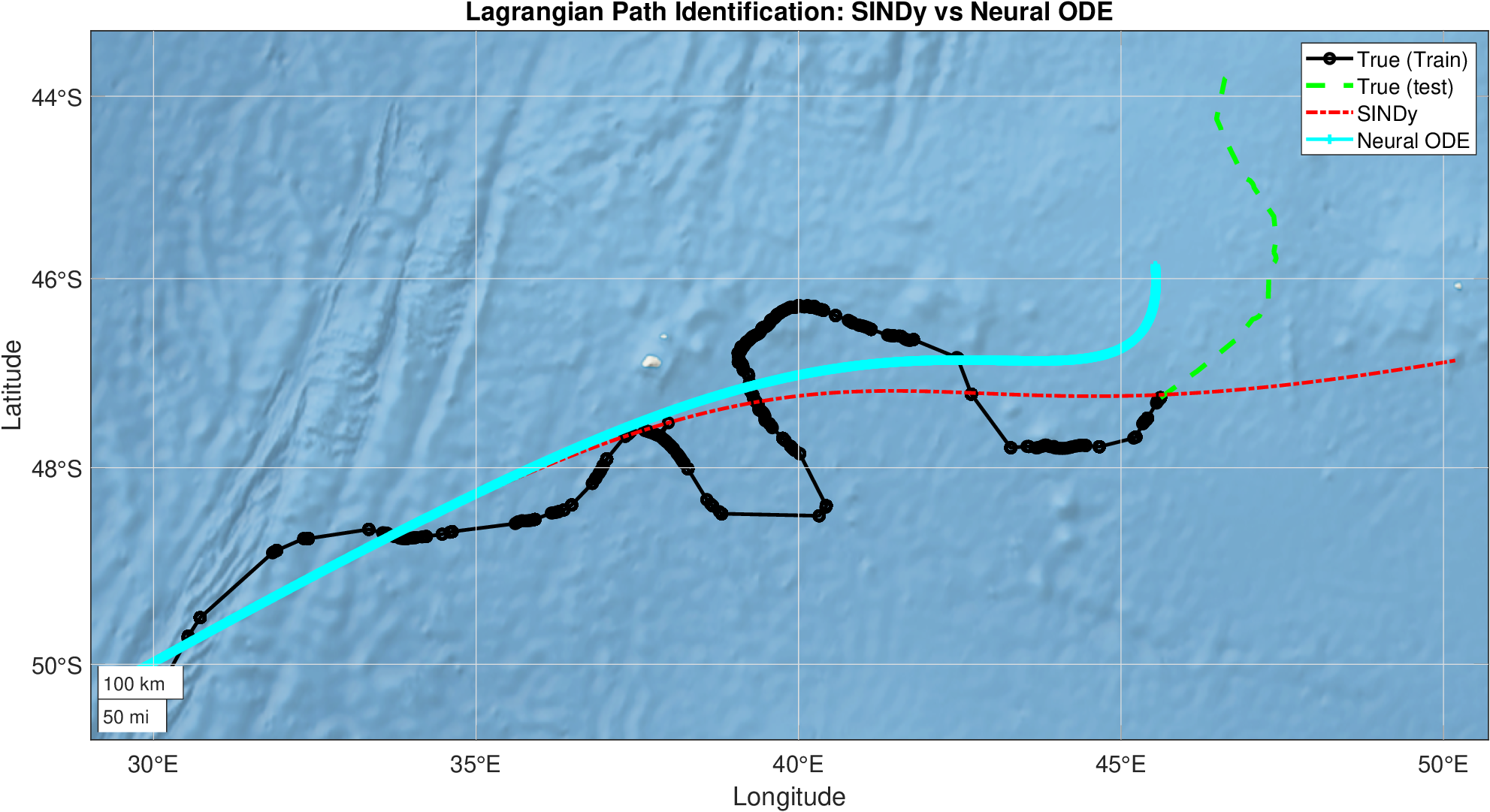}
\caption{\small The comparison of the SINDy using $d=5$ degree polynomials and no trigonometric terms vs neural ODE using $1500$ iterations in identification and prediction of the Drifter 1 (WMO$\#$ 1601747) trajectory.}
  \label{f:3}
\end{center}
\end{figure*}

As in the case depicted in Fig.~\ref{f:1}, the results depicted in Fig.~\ref{f:2} are obtained utilizing SINDy with $d=3$ degree polynomials and yet no trigonometric terms. However, the results obtained utilizing the neural ODE is for the case where the number of iterations is selected to be 1500. Although Drifter 1 trajectories retrieved by SINDY in the Fig.~\ref{f:1} and Fig.~\ref{f:2} are identical, Drifter 1 trajectories retrieved by the neural ODE are quite different, however both of them give quite successful results which can increase accuracy and the speed of the reconnaissance missions in the vast oceans and seas by reducing the possible search zones. The WGS84 distance between the endpoints of the actual trajectory vs the trajectory identified by SINDY with parameters given above is calculated as $4.4132\textnormal{x}10^5$ m as expected, whereas the distance between the endpoints of the actual trajectory vs the trajectory identified by neural ODE depicted in Fig.~\ref{f:2} is calculated to be $2.4719\textnormal{x}10^5$ m.

\begin{figure*}[htb!]
	\begin{center}
		\includegraphics[width=0.8\textwidth]{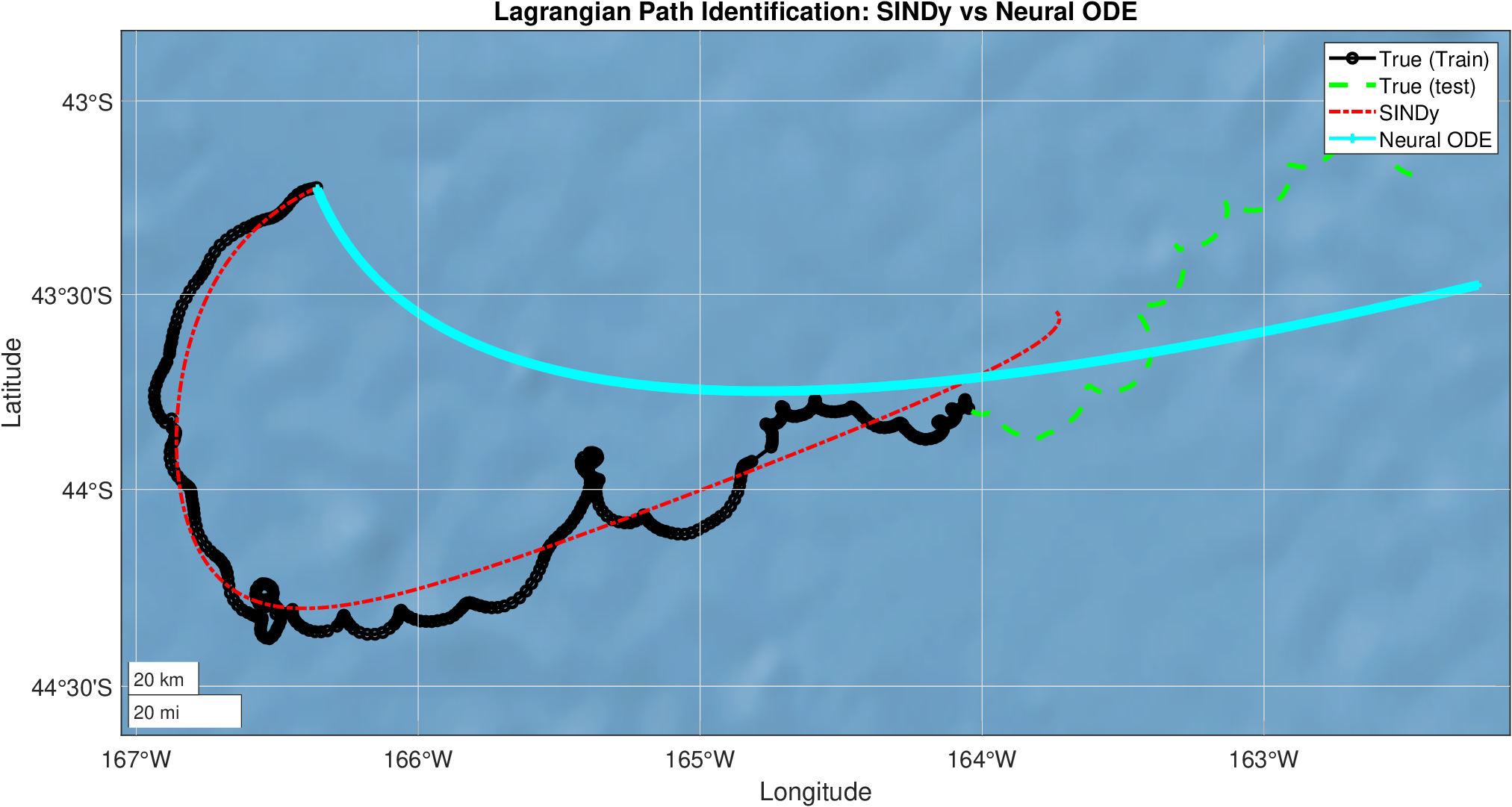}
	\caption{\small The comparison of the SINDy using $d=3$ degree polynomials and no trigonometric terms vs neural ODE using $1000$ iterations in identification and prediction of the Drifter 2 (WMO$\#$ 5501700) trajectory.}
	\label{f:4}
\end{center}
\end{figure*}

\begin{figure*}[htb!]
\begin{center}
   \includegraphics[width=0.8\textwidth]{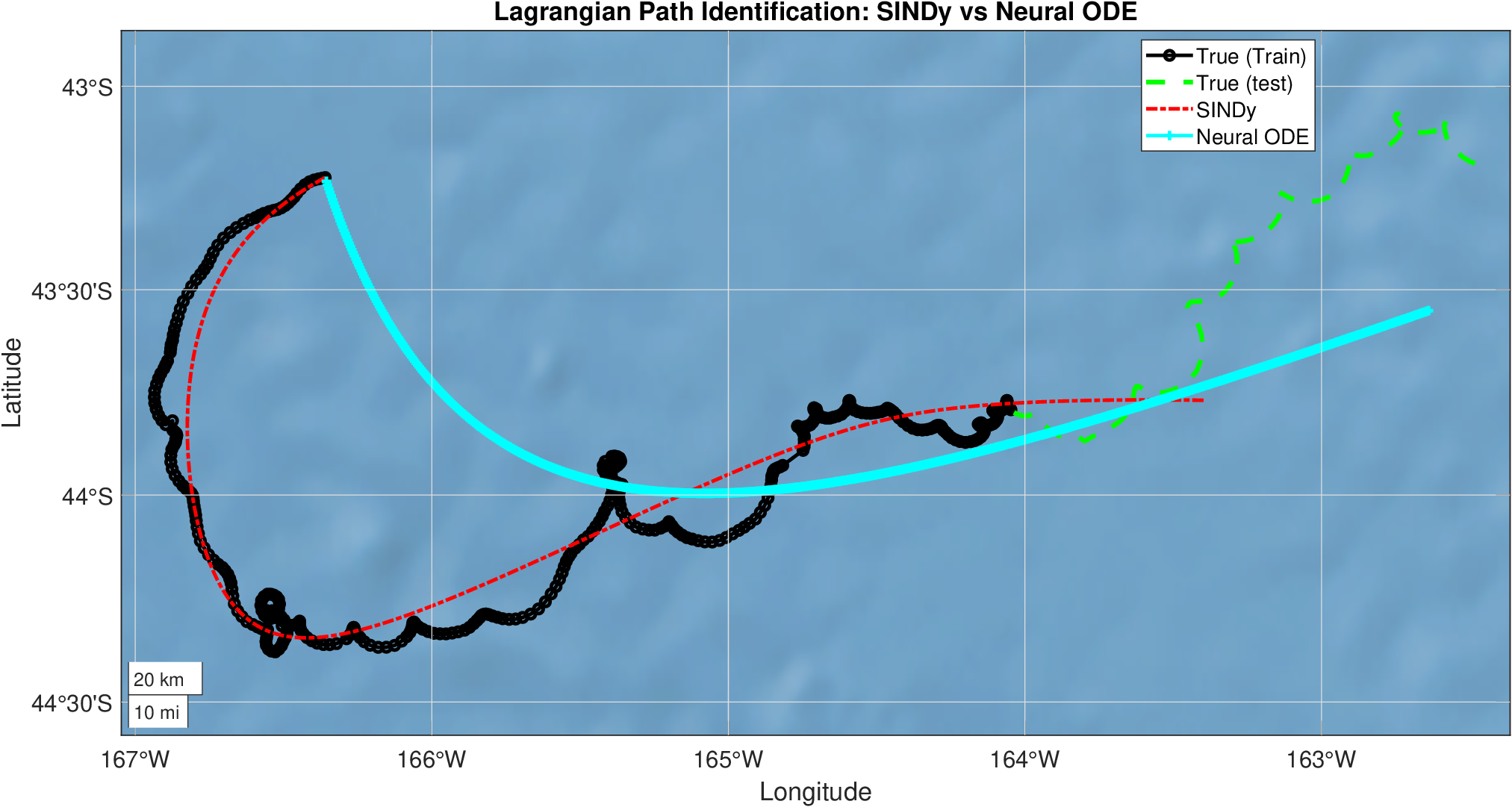}
\caption{\small The comparison of the SINDy using $d=3$ degree polynomials and with trigonometric terms vs neural ODE using $1000$ iterations in identification and prediction of the Drifter 2 (WMO$\#$ 5501700) trajectory.}
  \label{f:5}
\end{center}
\end{figure*}

Next, we turn our attention to assessing the performance of the SINDy algorithm when a higher order polynomial set is used without any trigonometric candidate function. In Fig.~\ref{f:3}, we depict the Drifter 1 path identification via SINDy where we have used $d=5$ order polynomials as candidate function and the number of iterations is set to be 1500 for neural ODE simulations as in the previous case. As the depicted trajectory, and the WGS84 distance between the endpoints of the actual trajectory vs the trajectory identified by SINDY calculated as $4.4132\textnormal{x}10^5$ m confirms, increasing the polynomial order from $d=3$ to $d=5$ does not lead to a significant change in the Lagrangian drifter path identification. Strikingly, although the computational parameters of the neural ODE are exactly the same as the case depicted in Fig.~\ref{f:2}, the identified trajectory in Fig.~\ref{f:3} is quite different.  The WGS84 distance between the endpoints of the actual Drifter 1 trajectory vs the same trajectory identified by neural ODE is calculated as $2.4493\textnormal{x}10^5$ m. Although the distance estimation by neural ODE seems to be more successful, the results of different runs of the neural ODE can be very different. In our simulations we also observe that not all of the runs give a convergent solution for the Lagrangian path identification and prediction. We observe that the divergence rate for the neural ODE is higher. Thus, both of the algorithms, especially the neural ODE must be used with care in making predictions about the dynamics of the Lagrangian drifters.

Next, we turn our attention to the second Lagrangian drifter trajectory, i.e. Drifter 2, data with WMO identification number of 5501700. This data set is strikingly more curved compared to the Drifter 1 data set, thus providing a good basis in assessing the validity of the two algorithms compared in this paper, as well as the effects of the trigonometric candidate functions of the SINDy algorithm. In Fig.~\ref{f:4}, we plot the identified and predicted trajectories of the Drifter 2 against its actual measurements. As before, the SINDy algorithm with $d=3$ order polynomials as candidate functions are used, the number of iterations of the neural ODE is selected to be 1000. 

For this case, the WGS84 distance between the endpoints of the actual trajectory vs the trajectory identified by SINDY with parameters given above is calculated as $1.0950\textnormal{x}10^5$ m, whereas the distance between the endpoints of the actual trajectory vs the trajectory identified by neural ODE depicted in Fig.~\ref{f:4} is calculated to be $3.6708\textnormal{x}10^4$ m.

Lastly, we investigate the effects of trigonometric candidate functions of the SINDy algorithm in retrieving the Drifter 2 data. We repeat the analysis depicted Fig.~\ref{f:4} with the same parameters, but the only difference is that we use trigonometric candidate functions in the SINDy algorithm. The WGS84 distance between the endpoints of the actual trajectory vs the trajectory identified by SINDY with parameters given above is calculated as $9.8306\textnormal{x}10^4$ m, whereas the distance between the endpoints of the actual trajectory vs the trajectory identified by neural ODE depicted in Fig.~\ref{f:5} is calculated to be $4.2022\textnormal{x}10^4$ m. 

A comparison of the results depicted in Figs.~\ref{f:4} and~\ref{f:5} reveals that although the neural ODE algorithm seems to exhibit a better performance in locating the last coordinates of drifter trajectory, deviation of its identified and predicted trajectory from the data seems to significantly larger from the actual measurements. Additionally, the divergence rate of the neural ODE resulting in no identified patterns is significantly higher than that of the SINDy algorithm. The results of the neural ODE seems to change significantly from run-to-run, at least for the parameters used and Lagrangian drifter trajectories identified. Another important lesson that can be learned form the comparison of Figs.~\ref{f:4} and~\ref{f:5} is that, the SINDy with trigonometric terms allows trajectories with higher curvatures to be identified more accurately, thus it is recommend to employ SINDY with its full potential by allowing for trigonometric candidate functions in retrieval of the Lagrangian drifter dynamics. 

The SINDy and neural ODE algorithms show promise in identifying the underlying dynamics of Lagrangian drifter paths as illustrated with our findings, however there are several avenues for improving our results. First, improving the quality of the input data has a significant effect of drifter path identification. The accuracy of the identified trajectories heavily relies on the resolution and reliability of the drifter trajectory location and velocity data. Implementing better data preprocessing approaches, such as noise reduction and outlier detection and removal, can lead to a more robust representation of the underlying Lagrangian drifter dynamics. Predetermined forcing in the marine environment by waves, currents and wind can be implemented as a forcing term in the SINDy algorithm. Additionally, integration of the complementary data sources such as oceanographic measurements or satellite imagery, could provide richer contextual information that helps refine the model identification.

Another potential improvement lies in the selection of library functions used in the SINDy algorithm. As shown in this paper, the performance of SINDy is sensitive to the choice of basis functions, which define the nonlinear terms in the dynamical model. Expanding the library of candidate functions to include a broader range of polynomial, trigonometric, or even custom-built basis functions tailored to specific oceanic conditions can enhance the algorithm's ability to capture complex dynamics. Furthermore, incorporating adaptive selection methods that automatically adjust the SINDy's library of candidate functions based on the characteristics of the data could lead to more accurate and representative models.

Finally, the integration of AI, DL techniques as well as the smart sensing techniques such as compressive sensing in the marine environment \cite{bayindirCAF} with SINDy and the neural ODE algorithms can also significantly enhance their performance and speed in identifying drifter paths. For instance, leveraging DL methods for feature extraction before applying SINDy and neural ODE could help uncover hidden patterns in the data that traditional approaches might miss. Moreover, combining SINDy and the neural ODE algorithms with reinforcement learning algorithms could allow for the optimization of model parameters. By fostering an interdisciplinary approach that merges traditional dynamical systems methods with cutting-edge machine learning (ML) techniques, researchers can significantly improve the identification and understanding of Lagrangian drifter dynamics.

\section{Conclusions}

In this article, we have examined the possible usage of the SINDy and the neural ODE algorithms in identification and prediction of the Lagrangian drifter trajectories. We showed that although both algorithms give quite acceptable results for the vast marine environment, the convergence of both of these algorithms is not guaranteed. However, SINDy algorithm is a more reliable tool due to its better consistency in identifying and predicting the Lagrangian drifter trajectories, at least for the data sets and parameters used in this paper. When the trigonometric candidate functions of the SINDY library are used, the results appear to be more successful in identifying the dynamics of trajectories with higher curvatures. Neural ODEs in their current form employed in this paper must be used with greater care, as the divergences of the path identifications are more commonly observed when they are used.

Lagrangian drifter path identification and prediction using SINDy and neural ODE algorithmic techniques has transformative potential across various fields in marine science. These methods leverage sophisticated algorithms to analyze data from oceanic drifters, enabling more accurate tracking of water movement and understanding of the governing rules of ocean dynamics. Possible benefits of these algorithmic retrieval techniques include but are not limited to enhanced modeling of ocean currents aids in studying marine ecosystems, better understanding of the species migration patterns and habitat connectivity, enhancing the efficiency of the climate models by offering insights into how ocean currents influence global weather patterns and climate change. Additionally, these algorithms can predict the movement of pollutants or debris in the ocean, facilitating quicker response efforts to environmental disasters and improving maritime safety. Real-time path predictions can guide rescue teams in locating lost vessels or individuals at sea, increasing the effectiveness of search efforts. 

\subsection*{Funding}This study is funded by the Turkish Academy of Sciences (TÜBA)-Outstanding Young Scientist Award Program (GEBİP), The Science Academy's Young Scientist Award Program (BAGEP) and the Research Fund of the İstanbul Technical University with project codes: MYL-2022-43642 and MDA-2023-45117.

\end{document}